\pdfoutput=1
\documentclass[aps,pra,reprint,amsmath,amssymb]{revtex4-1}
\usepackage{bm}
\usepackage{graphicx}
\usepackage[colorlinks]{hyperref}
\usepackage{mathrsfs}
\usepackage{times}

\begin{document}

\title{Quantum Scar States in Coupled Random Graph Models}

\author{Bhilahari Jeevanesan}
\email{Bhilahari.Jeevanesan@dlr.de}
\affiliation{Remote Sensing Technology Institute, German Aerospace Center DLR, 82234 Wessling, Germany}

\begin{abstract}
We analyze the Hilbert space connectivity of the $L$ site PXP-model by constructing the Hamiltonian matrices via a Gray code numbering of basis states. The matrices are all formed out of a single Hamiltonian-path backbone and entries on skew-diagonals.  Starting from this observation, we construct an ensemble of related Hamiltonians based on random graphs with tunable constraint degree and variable network topology. We study the entanglement structure of their energy eigenstates and find two classes of weakly-entangled mid-spectrum states. The first class contains scars that are approximate products of eigenstates of the subsystems. Their origin can be traced to the near-orthogonality of random vectors in high-dimensional spaces. The second class of scars has $\log 2$ entanglement entropy and is tied to the occurrence of special types of subgraphs. The latter states have some resemblance to the Lin-Motrunich $\sqrt{2}$-scars. 
\end{abstract}

\maketitle
\section{Introduction}
In the past few years a surprising kind of violation of {Srednicki's} strong eigenstate thermalization hypothesis (ETH) \cite{Srednicki1994} has been discovered.  It was found in experiments \cite{Bernien2017} that quantum simulators based on Rydberg atom arrays when quenched from certain initial density-wave patterns with periods $2$ and $3$ do not immediately relax toward thermal equilibrium, but instead show persistent coherent oscillations in the domain wall densities. 
A theoretical understanding of this behavior was provided by Turner et al. \cite{Turner2018} who start from the PXP model \cite{Lesanovsky2012} and identify special eigenstates that are related to the scar states studied by Heller \cite{Heller1984} in the 1980s. The authors of \cite{Turner2018} find among all the eigenstates a vanishingly small fraction of special states, the quantum many-body scars, which have area-law entanglement and are embedded in the middle of a thermal spectrum with volume-law entanglement.  The scar states are ultimately responsible for the periodic revivals seen in experiment. Quantum scars have since been found in a large number of other settings, ranging from spin models \cite{Moudgalya2018a, Moudgalya2018, Shiraishi2019, Schecter2019,Shibata2020, Surace2021, Langlett2022},  gauge theories \cite{Banerjee2021, Aramthottil2022}, fermionic and bosonic systems \cite{Kuno2020, Desaules2021, Su2023}  to quantum dimer models \cite{Biswas2022, Wildeboer2021}, for reviews see \cite{Regnault2022, Serbyn2021, Papic2022, Chandran2023}.  Scar states are also being explored for technological uses, for example for quantum sensing purposes \cite{Dooley2021} with the idea of exploiting the long coherence times seen in experiment. 

A natural question to ask is if such weak violations of ETH by scar states represent exceptions or whether the PXP model is just one in a large landscape of models displaying such behavior. Partial answers to this important question have been found in the literature by coming up with ways to construct mid-spectrum scar eigenstates. Spectrum-generating algebras (SGA) were employed in \cite{Moudgalya2020} in the context of the AKLT model to construct towers of non-thermal eigenstates with equidistant energy levels. SGAs and lattice scarring were discussed for open quantum systems in \cite{Buca2019}. A commutator framework was developed in \cite{Mark2020} and was used to give a short proof for the AKLT scar towers. Another elegant mechanism for scarring was provided by the Shiraishi-Mori construction \cite{Shiraishi2017} that uses local projection operators. For the PXP model itself, recently a probabilistic version was introduced in \cite{Surace2021}, where PXP basis states of Hamming distance one are connected randomly. The authors demonstrated the existence of `statistical scars'.
\begin{figure}[t] 
\centering{}\includegraphics[width=\columnwidth]{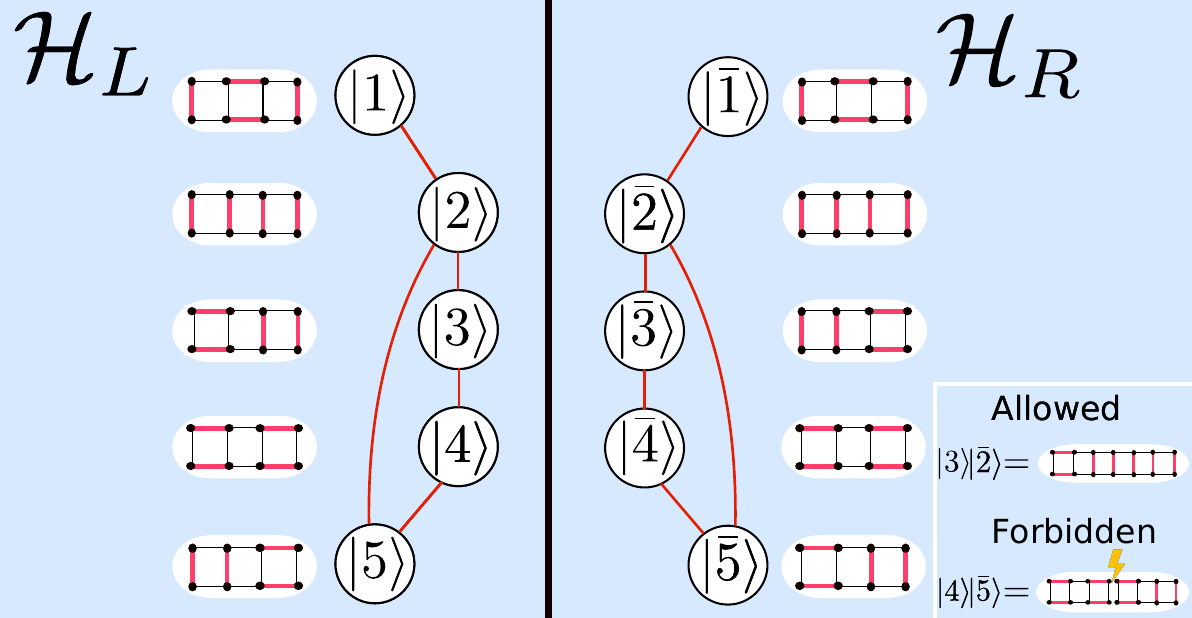}\caption{\label{FigGlue} An example of a constrained system that appears in the random graph ensemble constructed in this paper. Here a quantum dimer ladder with six plaquettes is split into a left and right half. A basis state $|i\rangle$ in the Hilbert-space $\mathcal H_L$ for the left part has a corresponding inverted basis state $|\bar i\rangle$ in $\mathcal H_R$ in the right part. The system is constrained in that not all combinations of basis states $|i\rangle\otimes|\bar j\rangle$ are permitted for the combined system, examples are shown in the corner. Other systems that lie in the random ensemble are the equivalent PXP model and the Rokhsar-Kivelson quantum dimer models.}
\end{figure}

In this work, we shed some light on the general question of the typicality of scars in constrained models. We extract some features from the PXP model and use them as a starting point to define the constrained Hamiltonians \eqref{eq:DefHamiltonian}. We specialize by choosing for $H_L$ and $H_R$  a collection of random graphs such that $H$ shares with the PXP model three properties. The first is the existence of  Hamiltonian-paths through the connectivity graph, which ensures that the Hilbert-space is connected. The second is graph-bipartiteness which leads to a symmetric energy spectrum. The third is the property of left-to-right inversion symmetry, which is present in the PXP model and leads there to an exponentially growing zero-energy subspace \cite{Turner2018, Turner2018a, Buijsman2022}. We show that our ensemble of Hamiltonians is generically quantum-ergodic and that it can harbor mid-spectrum scar states. We find two groups of scars with different origins and discuss the resemblance of the second group to the $\sqrt{2}$-scars discovered by Lin and Motrunich \cite{Lin2019}.

\section{The PXP Model in the Gray code basis}
Our starting point is the PXP Hamiltonian first introduced in \cite{Lesanovsky2012} in the context of an analog quantum simulator for Fibonacci anyons. It describes the Rydberg blockade physics of a collection of $L$ atoms arranged on a line. Each atom can be either in the ground state $\circ$ or the excited state $\bullet$. The blockade is realized as a prohibition that two neighboring sites must not both be excited, i.e. by forbidding configurations that contain the string $\bullet \bullet$ anywhere. The PXP Hamiltonian on $L$ sites is given by

\begin{equation}
H_L = \sum_{i=1}^{L} P_{i-1} X_i P_{i+1},
\label{eq:PXPH}
\end{equation}
here $X_i$ is the Pauli-X operator on site $i$ that interchanges $ {\circ \leftrightarrow \bullet}$. The $P_j$ are projection operators that assure the site $j$ is empty, i.e. $P_j = (1-Z_j)/2$, where $Z_j$ is the Pauli-Z operator on site $j$. We have set the coupling constant to unity.  In this paper we limit ourselves to the PXP-model with open boundary conditions (OBC), thus $P_0$ and $P_{L+1}$ are identity operators. The Hilbert space of the $L$-site PXP model is spanned by all strings of length $L$ composed of $\circ$ and $\bullet$ that do not violate the PXP constraint. With OBC the total number of such configurations is equal to the Fibonacci number $F_{L+2}$. 

\begin{figure}[t] 
\centering{}\includegraphics[width=0.9\columnwidth]{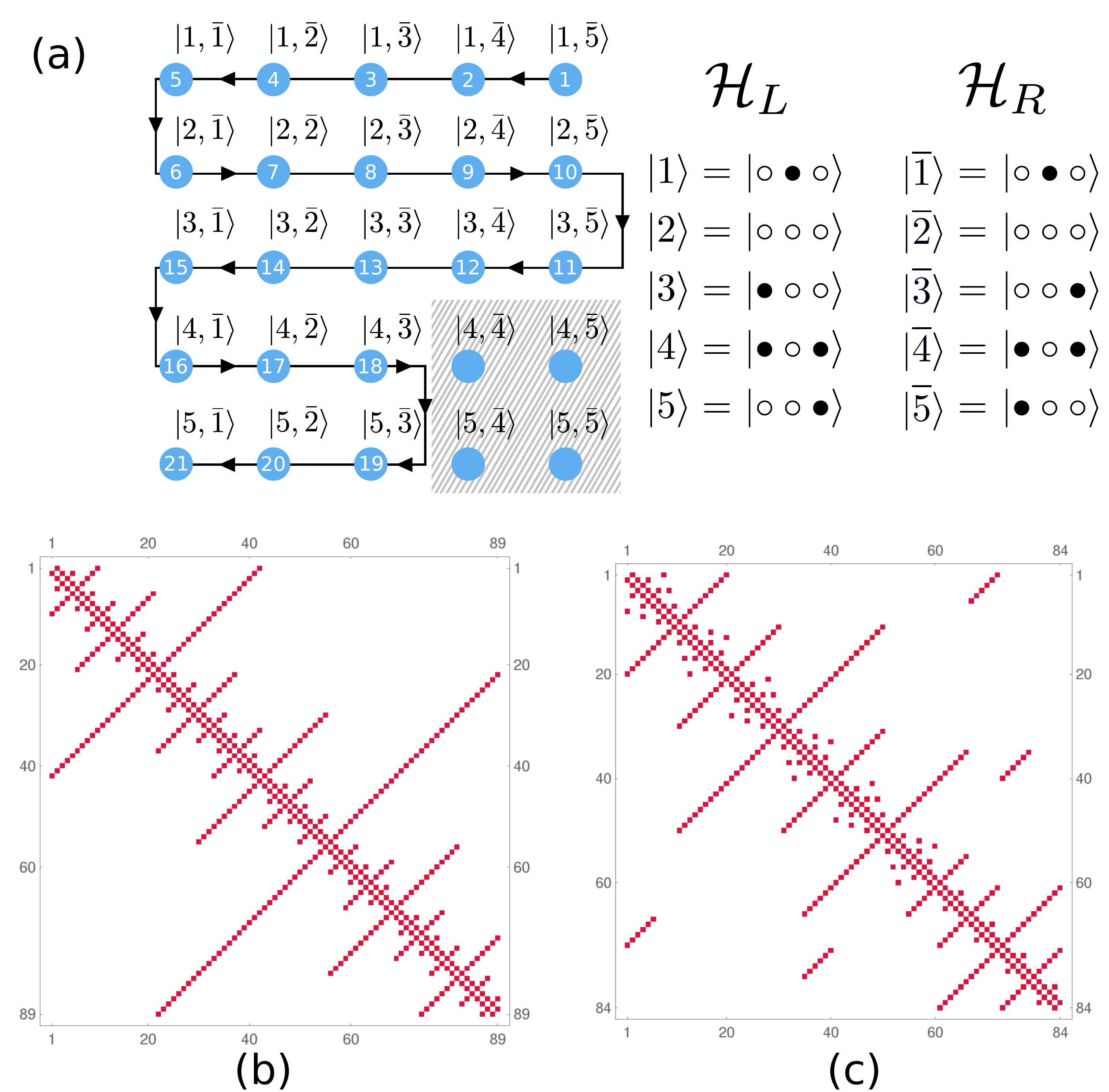}\caption{\label{Fig1} (a) The Hilbert space basis states of the PXP model for $L=6$ are composed of products of basis states for $L=3$. The hatched region shows the forbidden products. In our notation $\mathcal D = 5, \mathcal D' = 3$ and the number of basis states is given by the difference of the squares $5^2 - 2^2=21$, in general a PXP model with $2L$ sites has  $\mathcal D = F_{L+2}, \mathcal D' = F_{L+1}$, then the full dimension is $\mathcal D^2 - (\mathcal D-\mathcal D')^2 = F_{2L+2}$. (b) Adjacency matrix of the PXP model with $L=9$ sites in the Gray code basis. (c) Adjacency matrix of a random graph from our ensemble $\mathcal G (10, 6,0.5)$.}
\end{figure}

All matrix elements of the PXP Hamiltonian are either $0$ or $1$. This allows an interpretation in terms of a graph $\Gamma_L$ with adjacency matrix $H_L$ \cite{Turner2018}. The vertices of the graph are the computational basis states, i.e. all the $F_{L+2}$ configurations. Two vertices are connected by an edge if $H$ has a non-zero matrix element between the corresponding basis states. It was noticed in \cite{Turner2018a} that $\Gamma_L$ is well-known in the computer science literature as the Fibonacci-cube graph on $L$ vertices \cite{Hsu1993}. 

We will find it convenient to number the basis states since this allows us to recast the Hamiltonian \eqref{eq:PXPH} in an explicit matrix form. Different numberings of the basis states result in permuted rows and columns of the adjacency matrix, revealing different aspects of $H_L$.  An interesting property of $\Gamma_L$ is that it has a Hamiltonian path. This is a path through the graph that traverses all the vertices without repetition. In our first basis numbering we use an ordering of the basis states according to a Hamiltonian path. Another numbering relies on the Zeckendorf expansion \cite{Zeckendorf1972} and leads to an interesting recursion relation for the $H_L$ matrices, we discuss both bases in detail in Appendices \ref{sec:GrayCode} and  \ref{sec:Zeckendorf}.

An example of the Hamiltonian in the Gray code basis is shown in Fig. \ref{Fig1} (b).  An obvious property is the presence of the first super- and subdiagonal entries. This follows from ordering the basis states along a Hamiltonian path. This part of the Hamiltonian by itself corresponds to a simple nearest-neighbor tight-binding model. In addition to this backbone there are several entries arranged along skew-diagonals, see App. \ref{sec:GrayCode} for an explanation. We take these aspects of the PXP model as a starting point to construct a class of random graphs that contain the PXP model as one realization within an ensemble (namely for the choice $\mathcal D = F_{L+2}, \mathcal D' = F_{L+1}$, see below). 

\begin{figure*}[t] 
\centering{}\includegraphics[width=1.7\columnwidth]{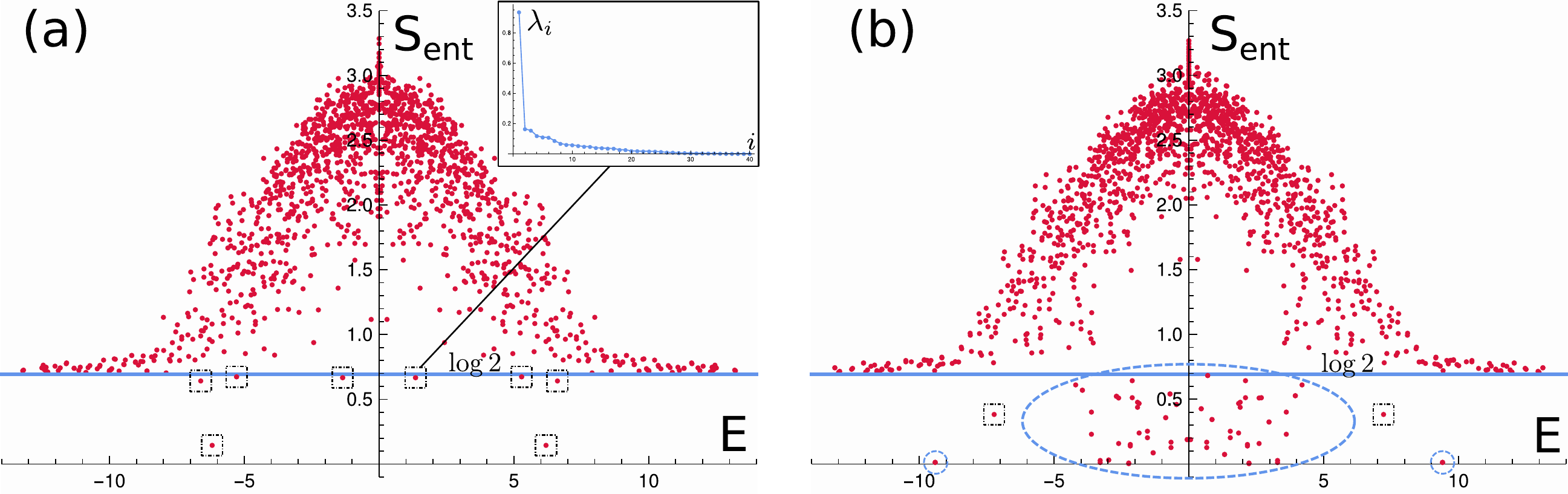}\caption{\label{FigEnt} Entanglement entropies of eigenstates resolved by their eigenenergies. (a) and (b) show the result for two Hamiltonians from the ensemble $\mathcal G(40, 34, 0.4)$. (a) The boxed eigenstates are scar states described by eq.  \eqref{eq:scargroup1}. (b) The circled scar states are described by eq. \eqref{eq:exactScarstri}-\eqref{eq:exactScarsdiam}. For this particular Hamiltonian, the scars originate from a $\triangleright$ subgraph in $H_L$. Note also the presence of the boxed scars.}
\end{figure*}
\section{The constrained Hamiltonian and random graph realizations}
Let us take a system (the spatial dimensionality is unimportant for the construction) with an orthogonal basis $\left\{|1\rangle,|2\rangle,\dots,|\mathcal D\rangle\right\}$. In general, the Hilbert space dimension $\mathcal D$ will be exponential in the system size. Next, we double the volume of the system by joining a left-to-right inverted copy to itself. The Hilbert spaces of the left and right parts we denote respectively by $\mathcal H_L$ and $\mathcal H _R$. We label the state in $\mathcal H_R$ that  corresponds to $|i\rangle$ in $\mathcal H_L$ by $|\bar i\rangle$, see Fig. \ref{FigGlue} for an example involving the quantum dimer ladder. The total system has basis states 
$|i\rangle | \bar j\rangle$, thus the dimensionality of the joint system can be as large as $\mathcal D^2$, but we assume that the system is constrained in such a way that not all states $|i\rangle$ and $|\bar j\rangle$ are compatible with each other. We introduce a cutoff $\mathcal D'$ such that a state $|i\rangle |\bar j \rangle$ is prohibited if $i$ and $j$ are simultaneously larger than $\mathcal D'$, otherwise it is allowed, see Fig. \ref{Fig1} for an example. Then the Hilbert space dimension of the full problem is $\mathcal D^2 - (\mathcal D-\mathcal D')^2$ and the Hamiltonian is given by
\begin{eqnarray}
\label{eq:DefHamiltonian}
H &=& \Pi \left[H_L \otimes \mathbb{I}_R  + \mathbb{I}_L \otimes  H_R\right] \Pi\\
H_L&=& \sum_{i,j=1}^\mathcal D h_{ij} |i\rangle \langle j| , \ H_R = \sum_{i,j=1}^\mathcal D h_{ij} |\bar i\rangle \langle \bar j|,
\end{eqnarray}
where  $\mathbb{I}_L =  \sum_i |i \rangle \langle i |$, $\mathbb{I}_R =  \sum_j |\bar j \rangle \langle \bar j |$ and $h_{ij} = h_{ji}$ is a matrix with entries $0$ or $1$. The Hamiltonians $H_L$ and $H_R$ describe the left and right subsystems. The projection operator $\Pi$ assures that the subsystems are never in incompatible states.  Formally, we can write it as $\Pi =\sum_c |c\rangle \langle c|$, where the sum is over all $|c\rangle = |i\rangle |\bar j\rangle$ with $i$ and $j$ not simultaneously larger than $\mathcal D'$.  Thus $\Pi$ is a projector that generalizes the $P$ operator appearing in the PXP Hamiltonian. Like $P$ in the PXP model, the projector $\Pi$ is ultimately responsible for rendering \eqref{eq:DefHamiltonian} non-trivial, since without it the left and right parts of the system would not interact. The eigenstates of $H$ would then just be the products $|\varphi_L\rangle |\varphi_R\rangle$ with the factors being the eigenstates of $H_L$ and $H_R$ respectively. When $\Pi$ is present such a product is generally not an eigenstate. Nevertheless, there are some near-eigenstates that lead to scarring, as we discuss below. The constraint is tunable by varying $\mathcal D'$ and at the point $\mathcal D' = \mathcal D$ the system is non-interacting since $\Pi$ becomes the identity operator. A different approach was introduced in \cite{Langlett2022}, where instead of our projector $\Pi$ the authors use an explicit interaction term in the Hamiltonian to couple the left and right parts. 

Since $h_{ij}$ has only $0$ and $1$ entries, the Hamiltonians $H_L$ and $H_R$ can be viewed as adjacency matrices of graphs. For this reason, we will sometimes refer to $H_L, H_R$ and $H$ as graphs. 

Clearly, the PXP model with an even number of sites falls into the class \eqref{eq:DefHamiltonian} of models: If we take for $H_L$ the PXP Hamiltonian of size $L$ and choose for $\Pi$ an operator that forbids state pairs $|i\rangle |\bar j \rangle$ where excited sites would be bordering the joining interface, then $\mathcal D' = F_{L+1}$ and $H$ is the PXP Hamiltonian $H_{2L}$. Another example of a system of type \eqref{eq:DefHamiltonian} is the {\it two-dimensional} quantum dimer model of Rokhsar and Kivelson on the square lattice \cite{Rokhsar1988}. Here the $\Pi$ operator projects out those states that would violate the `one dimer per site' constraint at the joining interface, see Fig. \ref{FigGlue}. Incidentally, the {\it one-dimensional} dimer ladder, cast in the language of hard-core bosons, was shown in \cite{Chepiga2019} to be equivalent to the PXP model, thus the PXP model and quantum dimer models are intimately related.

The inversion operator $\mathcal I$ mirror-reverses the left and right systems, thus it maps the state $|i\rangle|\bar j\rangle$ to $|j\rangle|\bar i\rangle$ and is therefore given by
\begin{equation}
\mathcal I = \sum_{i,j} |j\rangle|\bar i\rangle \langle i|\langle \bar j|
\end{equation}
and by construction we have $\mathcal I H \mathcal I = H$. Hence all the non-degenerate eigenstates of $H$ are also eigenstates of $\mathcal I$ with quantum numbers $+1$ or $-1$.  In carrying out the level-statistics analysis below, we always remove one of the sectors beforehand. 

In the following, we generate $H_L$, and thereby the left-to-right inverted $H_R$,  and study the resulting Hamiltonian $H$ of \eqref{eq:DefHamiltonian} by performing exact numerical diagonalization. We choose $H_L$ such that it has as a backbone the Hamiltonian path ${|1\rangle \leftrightarrow |2\rangle  \leftrightarrow \dots  \leftrightarrow |\mathcal D\rangle}$. In this way, we assure that the Hilbert space does not split into disconnected subspaces. Because $H_L$ and $H_R$ have Hamiltonian paths, $H$ itself has one, see Fig. \ref{Fig1} (a) for the construction. In the spirit of the PXP model, we let $H_L$ be the adjacency matrix of a graph with $\mathcal D$ vertices. We choose the graph randomly by adopting the probabilistic ensemble approach, pioneered by Erd\H{o}s and R\'enyi in their seminal paper \cite{Erdoes1960} from 1960, to our situation. Concretely this means that starting from the backbone, we consider each pair of vertices exactly once and insert an edge between them with a probability $p$. In order to retain the bipartiteness symmetry introduced by the backbone, we only consider adding edges that do not violate this symmetry. As a consequence, every eigenstate of $H$ with energy $E$ has a partner state with energy $-E$, see App. \ref{app:Bip}. To summarize, our model is fully characterized by the parameters $\mathcal D$, $\mathcal D'$ and the probability $p$. We denote the ensemble of graphs generated with this set of parameters by $\mathcal G(\mathcal D,\mathcal D', p)$. By carrying out a level statistics analysis of the exact eigenstates of the model we find that for $\mathcal D' < \mathcal D$ and for generic $p$ the Hamiltonians $H$ are quantum-ergodic, see App. \ref{sec:LevelStat} for details. 

In the following, we study the entanglement entropy of eigenstates of Hamiltonians from the ensemble ${\mathcal G(40, 34, p)}$ for various values of $p$. Let $|\psi \rangle$ be a non-degenerate eigenstate with the decomposition
\begin{equation}
|\psi\rangle = \sum_{i,j=1}^{\mathcal D} \psi_{ij} |i\rangle |\bar j \rangle.
\end{equation}
Since the inversion operator commutes with the Hamiltonian, we have $\mathcal I |\psi\rangle = \pm |\psi \rangle$. Moreover, $\mathcal I |i\rangle |\bar j \rangle = |j\rangle |\bar i \rangle$ implies $\psi_{i j} = \pm \psi_{j i}$, i.e. $\psi$ is symmetric if it has inversion quantum number $+1$ and anti-symmetric if the quantum number is $-1$. Let us denote the eigenvalues of $\psi_{ij}$ by $\lambda_m$. The eigenvalues of the reduced density matrix of the left system, $\rho_L=\text{Tr}_R (|\psi\rangle\rangle \langle |\psi|) $, are given by the Schmidt values $|\lambda_m|^2$ with normalization implying $\sum_{m=1}^{\mathcal D} |\lambda_m|^2 = 1$. The bipartite von Neumann entanglement entropy between the left and right parts in the state $|\psi\rangle$ is $S_\text{ent} = -\sum_{m=1}^{\mathcal D} |\lambda_m|^2 \log|\lambda_m|^2$.

The Fig. \ref{FigEnt} (a) shows a representative plot of the entanglement entropies of eigenstates versus their energies from the $\mathcal G(40,34,0.4)$ ensemble with matrices of size ${1564\times1564}$. We find typically that almost all eigenstates have entanglement entropy larger than $\log 2$. Yet quite frequently there appear a small number of states, the boxed points in Fig.  \ref{FigEnt}, with entanglement entropy lower than this. Inspection of the corresponding matrix $\psi_{ij}$ revealed in all cases that it has a dominant eigenvalue with $\lambda_1^2 \gtrsim 0.8$, while the other eigenvalues are much smaller, see the inset in Fig. \ref{FigEnt} (a) for an illustration. Denoting by $\bm \psi$ the corresponding normalized eigenstate, we can write this scar state as the rank-one approximation
\begin{equation}
\label{eq:scargroup1}
\psi_{ij} \approx \lambda_1 \psi_i \psi_j
\end{equation}
with a sub-thermal entanglement entropy $S_\text{ent} \approx  -\lambda_1^2 \log \lambda_1^2 $. 
Remarkably the eigenstate $\bm \psi$ is also nearly an eigenstate of $H_L$. In fact, we find that in this ensemble there is always an eigenstate $\bm \varphi$ of $H_L$ that has a large overlap ${|\bm \varphi \cdot \bm \psi| \sim 0.9}$. At $\mathcal D' = \mathcal D$, a product state with $\psi_{ij} = \varphi_i \varphi_j$ is always an exact eigenstate of $H$ if $\bm \varphi$ is an eigenstate of $H_L$. As $\mathcal D'$ is lowered this is generally not so. Yet what we numerically show in App. \ref{app:overlap} is that one can still always find at least one $\bm \varphi$ such that the state $\psi_{ij} = \varphi_i \varphi_j$ is close to an eigenstate of $H$. We study the average overlap between the latter states in App. \ref{app:overlap} and find that it only decreases slowly with decreasing $\mathcal D'$. Heuristically, this phenomenon can be traced back to the fact that two random vectors in a high-dimensional space are nearly orthogonal to each other with large probability, see App. \ref{app:overlap}. Thus the near-product-state form in eq. \eqref{eq:scargroup1} holds for a few eigenstates of $H$ and is the reason for the low entanglement of the boxed states. 

\begin{figure}[t] 
\centering{}\includegraphics[width=0.7\columnwidth]{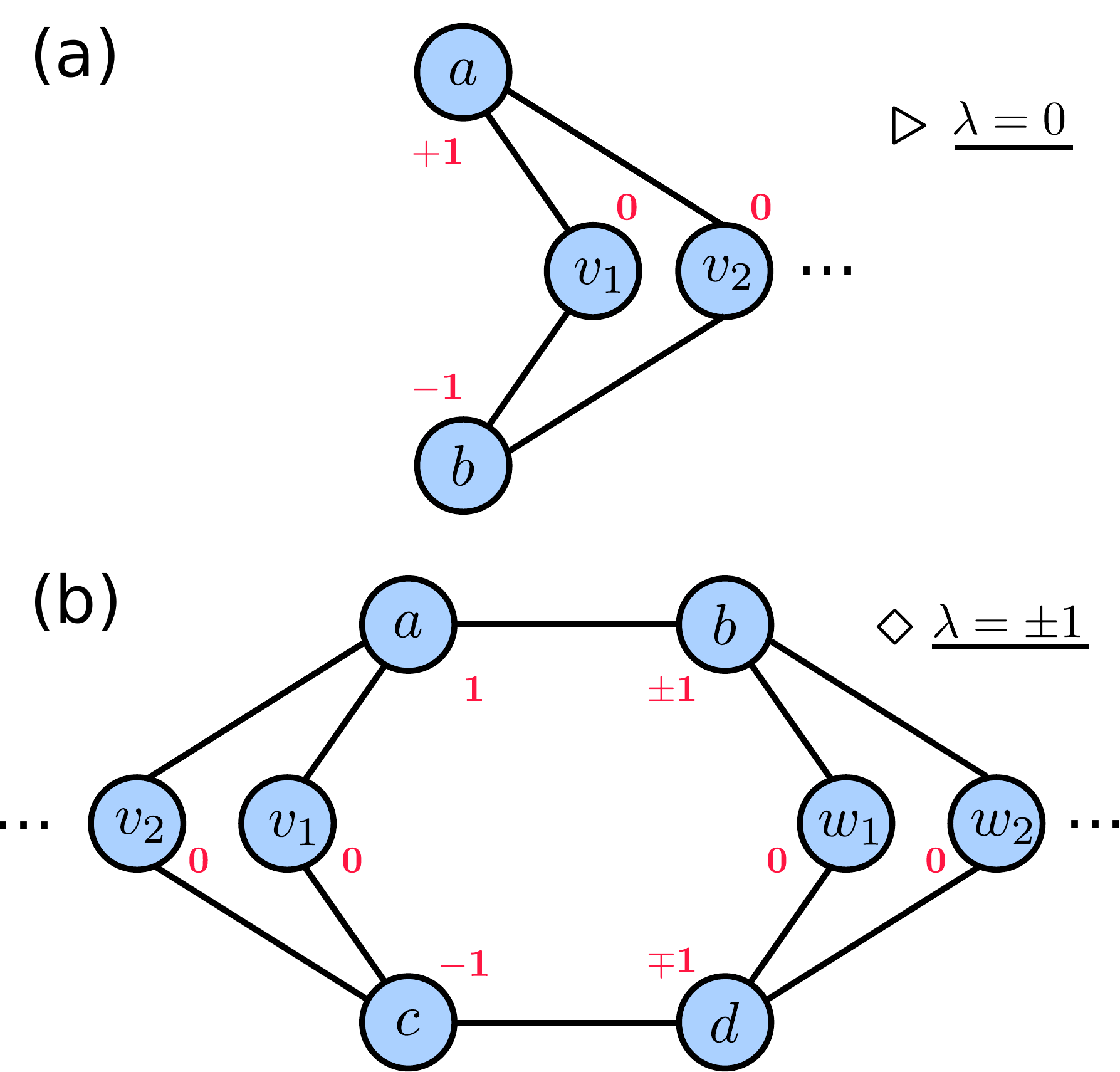}\caption{\label{Fig:Eigen}The appearance of either of these two types of subgraphs in $H_L$ generates sparse eigenvectors, denoted $\psi^\triangleright$ and $\psi^\diamond_\pm$, with eigenvalues $0$ and $\pm 1$ respectively, see \eqref{eq:psitriangle} and \eqref{eq:psidiamond}. In (b) all the $v$ and $w$ vertices may be connected in an arbitrary way amongst each other, without affecting the eigenstate property. Shown in red on any vertex $n$ is the $n^\text{th}$ entry of the respective eigenvector.}
\end{figure}

Apart from the boxed scar states, there appears for a certain fraction of random graphs in the ensemble, a large number of eigenstates near the middle of the spectrum that have double degeneracy and less than $\log 2$ entanglement entropy, these are the circled states in Fig. \ref{FigEnt} (b). Analysis reveals that the states have a simple entanglement spectrum: Their reduced density-matrices $\rho_L$ have rank two, i.e. all eigenvalues of $\rho_L$ are identically zero, except for two. Moreover, we find empirically that the appearance of the special states is always accompanied by the appearance of either of two kinds of subgraphs of $H_L$ with sparse eigenstates, see Fig. \ref{Fig:Eigen}. We note that the `statistical scars' of \cite{Surace2021} were also linked to the appearance of special subgraphs. Yet the details of how the scars in our model emerge from the subgraphs are quite different, in particular the scars seen in Fig. \ref{FigEnt} (b) do not appear only at energies $0, \pm 1$ or $\pm \sqrt{2}$, but occur for generic energy values.  The following explains all these observations.

The first type of sparse eigenstate of $H_L$ has zero energy. It corresponds to the situation shown in Fig. \ref{Fig:Eigen} (a). There are two special vertices $a$ and $b$ that are not connected to each other, but whenever a third vertex $c$ is connected to $a$ it is also connected to $b$ and vice versa. Thinking in terms of $H_L$ in matrix form, this implies that the column of vertex $a$ and the column of vertex $b$ are identical. Thus the state 
\begin{equation}
\label{eq:psitriangle}
|\psi^\triangleright\rangle \equiv \frac{|a\rangle - |b\rangle}{\sqrt{2}}
\end{equation}
is an eigenstate of $H_L$ with energy $0$. 

Another type of special eigenvector with eigenvalue $\pm 1$ occurs whenever the graph of $H_L$ contains the subgraph shown in Fig. \ref{Fig:Eigen} (b). Here there are four special vertices $a,b,c,d$ such that $a$ is not connected to $\{c,d\}$ and $b$ is not connected to $\{c,d\}$. Moreover, $a$ connects to $b$ and $c$ connects to $d$.  Other vertices $v_i$ are either connected to both $a$ and $c$ or not connected to either. Similarly, a vertex $w_i$ is either connected to both $b$ and $d$ or to neither, see Fig. \ref{Fig:Eigen} (b). With such a network structure the two states
\begin{equation}
\label{eq:psidiamond}
|\psi^{\diamond}_\sigma \rangle \equiv \frac{|a\rangle + \sigma  |b\rangle - |c\rangle -\sigma  |d\rangle}{2}
\end{equation}
with $\sigma = \pm 1$ are always eigenvectors of $H_L$. Their  eigenenergies are $\sigma = \pm 1$. We give the detailed proof in App. \ref{app:EigProof}.
In all the instances where we found a multitude of states with small entanglement, as in Fig. \ref{FigEnt} (b),  we always found  that $|\psi^\triangleright\rangle$ and $|\psi^\diamond\rangle$ had
\begin{eqnarray}
\label{eq:condition}
a,b \leq \mathcal D' \text{ and } a,b,c,d \leq \mathcal D'
\end{eqnarray}
for the respective situations. 

Starting with the states \eqref{eq:psitriangle}-\eqref{eq:psidiamond} and the condition \eqref{eq:condition}, we can construct exact eigenstates of $H$ with low entanglement. Let $|\psi_n\rangle$ and $|\bar \psi_n\rangle$ be eigenstates of $H_L$ and $H_R$ with energy $E_n \neq 0 , \pm 1$. Then the latter states are orthogonal to $|\psi^\triangleright\rangle$ and $|\psi^\diamond\rangle$. The vectors
\begin{eqnarray}
\label{eq:exactScarstri}
|\psi^{\triangleright \text{-scar}}_n \rangle&\equiv& \frac{|\psi_n\rangle \otimes |\bar \psi^{\triangleright}\rangle  \pm  |\psi^{\triangleright}\rangle \otimes |\bar \psi_n\rangle}{\sqrt{2}} \\
\label{eq:exactScarsdiam}
|\psi^{\diamond\text{-scar}}_{n,\sigma} \rangle&\equiv& \frac{|\psi_n\rangle \otimes |\bar \psi_\sigma^{\diamond}\rangle  \pm  |\psi_\sigma^{\diamond}\rangle \otimes |\bar \psi_n\rangle }{\sqrt{2}} 
\end{eqnarray}
are normalized eigenstates of $H$ with definite inversion quantum number $\pm 1$, depending on which sign is chosen on the right hand side. The reason why these are exact is that because of \eqref{eq:condition}, the states $|\psi^{\triangleright}\rangle, |\psi^{\diamond}_\sigma\rangle$  are formed out of basis states that can never be incompatible with their partnering factors in the products in \eqref{eq:exactScarstri}-\eqref{eq:exactScarsdiam} and thus  are never affected by the $\Pi$ operator in the full Hamiltonian in eq. \eqref{eq:DefHamiltonian}. 

The eigenvalue of the state in eq. \eqref{eq:exactScarstri} is $E_n$, the eigenvalue of the two states in eq. \eqref{eq:exactScarsdiam} is $E_n+\sigma$. Since there are roughly $\mathcal D$ choices for $|\psi_n\rangle$, there are about $2\mathcal D$ states of the form \eqref{eq:exactScarstri} or \eqref{eq:exactScarsdiam}. Due to the inversion quantum number all three states in \eqref{eq:exactScarstri}-\eqref{eq:exactScarsdiam} are doubly degenerate, explaining our numerical observation. Since $\langle \psi^\triangleright|\psi_n\rangle = 0$, it follows from eq. \eqref{eq:exactScarstri} that the density matrix has only two non-zero eigenvalues $1/2$ and $1/2$, yielding a sub-thermal entanglement entropy of $S_\text{ent}^\triangleright = \log 2$. Similarly for $E_n\neq \pm 1$ we have $\langle \psi_\pm^\diamond|\psi_n\rangle = 0$ and from eq. \eqref{eq:exactScarsdiam} we infer $S_\text{ent}^\diamond= \log 2$. The reason why the entropies in Fig. \ref{FigEnt} (b) are lower is that due to the double-degeneracy the exact diagonalization yields arbitrary superpositions.

In the App. \ref{app:probdist} we study the probability with which the $\triangleright$ and $\diamond$ subgraphs appear when $H_L$ is randomly generated with edge-insertion probability $p_\text{edge}$. We find somewhat expectedly that the $\triangleright$ state is much more frequent than the $\diamond$ states. Scarring due to $\triangleright$ and $\diamond$ states becomes increasingly rare as $\mathcal D'$ is decreased because it becomes less likely that the conditions $\eqref{eq:condition}$ can be met.

As shown in \cite{Lin2019}, for the PXP model with $2L$ sites ($L$ even), the reduced bipartite density-matrix in the $\sqrt{2}$-scar state also has exactly two non-zero eigenvalues, $1/2$ and $1/2$, resembling the states we just found. By numerically inspecting the $\sqrt{2}$-scars in the PXP model with $2L$ sites further, we find that they are constructed from states of the $L$ site model, similar to eq. \eqref{eq:exactScarstri} and eq. \eqref{eq:exactScarsdiam}:
\begin{eqnarray}
\label{eq:exactScarLMi}
|\psi^{\sqrt{2}\text{-scar}}_{2L} \rangle =  \frac{|\psi^0 \rangle \otimes |\psi^{\sqrt{2}\text{-scar}}_{L} \rangle  -  |\psi^{\sqrt{2}\text{-scar}}_{L} \rangle \otimes |\psi^0 \rangle}{\sqrt{2}}
\end{eqnarray}
where $L$ is even and $|\psi^0 \rangle$ is a specific zero-energy eigenstate. Interestingly, each product in the numerator of eq. \eqref{eq:exactScarLMi} by itself has some terms with two $\bullet$ next to each other, violating the PXP constraint. However, by taking the difference such terms drop out. 

\section{Summary and Outlook}We have proposed an ensemble of random graphs whose adjacency matrices are quantum-ergodic Hamiltonians and have shown that they can have two types of scar states embedded in the spectrum. For the construction it was crucial that the Hamiltonians in eq. \eqref{eq:DefHamiltonian} have exclusively the matrix elements $0$ or $1$, i.e. that they are binary matrices. Only this assumption led to the interpretation of $H$ as a graph. This in turn led to the possible presence of special subgraphs with sparse eigenstates that give rise to scar states.  The PXP Hamiltonian is a binary matrix, moreover there are other constrained systems, such as quantum dimer models, that are also of this form. Thus they appear in our ensemble of graphs for particular choices of $(\mathcal D, \mathcal D')$. Therefore the Hamiltonian eq. \eqref{eq:DefHamiltonian} opens up an interesting direction for future explorations.  

The graph interpretation of the Hamiltonians in eq. \eqref{eq:DefHamiltonian} also provides a direct connection to experiments. In this picture, the many-body problem eq. \eqref{eq:DefHamiltonian} is translated into a tight-binding model for a single particle on the corresponding graph. The system, when initialized in one of its basis states, will carry out a quantum random walk on the graph. Thermalization corresponds to the immediate spreading of the amplitude into the rest of the graph. For scarred eigenstates, there will instead be a coherent periodic refocusing to the initial state by quantum interference. 
Graphs have played an important role in recent advances in noisy intermediate-scale quantum devices \cite{Preskill2018, Deng2023, Harrigan2021, Byun2022} and other quantum technologies \cite{Ebadi2022, Nguyen2023,Breuckmann2021, Suprano2022}. Quantum random walks have been realized in several experiments, in particular on photonic platforms for a multitude of different graph topologies \cite{Caruso2016, Nejadsattari2019, 	Perets2008,Bian2017,AspuruGuzik2012, Adao2022}. It is therefore an interesting question whether the quantum scars found in the present work can be observed on such platforms.

\acknowledgements{It is a pleasure to thank Debasish Banerjee, Sergej Moroz and Arnab Sen for stimulating discussions. 

\clearpage 
\appendix
\begin{widetext}
\section{Graph Bipartiteness and symmetry of the spectrum}
\label{app:Bip}
The random graphs that we generate in the main text are all bipartite by construction, thus we can assign each vertex to be of A-type or B-type. A vertex of one type only connects to a vertex of the other type, but never to one of the same type. We can now show that as a consequence of this property, the eigenvalue spectrum of $H$ is symmetric. Let $Q$ denote the diagonal matrix with $Q_{ii}=1$ if $i$ is an A-type vertex and $Q_{ii}=-1$ if $i$ is of B-type. If the matrix $Q$ is multiplied with a vector $\psi$, it changes the sign of the $B$-vertex entries of $\psi$ . Consider
the eigenvalue equation
\[
H\psi=\epsilon\psi.
\]
Multiplying this by $Q$ and using $Q^{2}=1$, we obtain
\[
QHQ\ Q\psi=\epsilon Q\psi.
\]
Since $H$ is bipartite  it follows that in order for 
\[
\left(QHQ\right)_{ij}=Q_{ii}H_{ij}Q_{jj}
\]
to be non-zero, we must have that if $i$ is A-type then $j$ has to be B-type and vice versa. Thus $Q_{ii}Q_{jj}=-1$ and we obtain $QHQ=-H$, hence
\[
HQ\psi=-\epsilon Q\psi.
\]
Clearly, $Q\psi$ is an eigenvector with eigenvalue $-\epsilon$ and the full spectrum consists of eigenvalues that come in $\pm$ pairs. 

\section{Entanglement entropy of $|\psi\rangle$  and $Q|\psi\rangle$}
\label{app:QEntanglement}
Given a state $|\psi\rangle$ in the Hilbert space of one of our ensembles, we prove that it has the same entanglement spectrum as the state $Q|\psi\rangle$. To show this, we decompose $|\psi \rangle$ into a sum over the basis states $s$, which in turn we factor into a left and right part:
\begin{equation}
|\psi\rangle  = \sum_{s} \psi_s |s\rangle = \sum_{(m,n)} \psi_{m n } |a_m\rangle |b_n\rangle
\end{equation}
We have shown in Fig. \ref{Fig1}(a) that a Hamiltonian path through the system always exists. This path helps to uniquely assign each basis state an $A$ or $B$ label. But note in the same diagram that within each row the first factor is fixed. Thus we can assign each factor individually an $A$ or $B$ label. We do this by associating with each factor a sign $\sigma_m, \sigma_n \in \{-1,+1\}$, such that the sign of their product determines the type of $|a_m\rangle |b_n\rangle$. 
Acting with $Q$ on this state yields
\begin{equation}
Q|\psi\rangle  = \sum_s \psi_{m n } \sigma_m \sigma_n |a_m\rangle |b_n\rangle.
\end{equation}
Clearly, we can absorb the signs by redefining the basis of each factor $|\tilde a_m\rangle = \sigma_m |a_m\rangle $ and $|\tilde b_n\rangle =  \sigma_n |b_n\rangle$. Such a local basis change does not affect the entanglement spectrum, since the Schmidt values of $\psi_{mn}$ remain invariant under orthogonal basis transformations of the left and right Hilbert spaces separately. Thus we conclude that $|\psi\rangle$ and $Q|\psi\rangle$  have reduced density matrices with identical Schmidt values and therefore identical entanglement entropies. 

\section{Eigenstates of $\triangleright$ and $\diamond$ subgraphs}
\label{app:EigProof}
Here we write out explicit proofs that the two subgraphs do indeed have the claimed eigenstates. Starting with 
\begin{equation}
|\psi^\triangleright\rangle \equiv \frac{|a\rangle - |b\rangle}{\sqrt{2}}
\end{equation}
we consider the action of $H_L$ on it:
\begin{equation}
H_L |\psi^\triangleright\rangle  = \sum_{ij} h_{ij} |i\rangle| \langle j| \frac{|a\rangle - |b\rangle}{\sqrt{2}} = \sum_{i}  \frac{h_{ia}- h_{ib}}{\sqrt{2}} |i\rangle  
\end{equation}
Now if $i$ is a vertex in the $\triangleright$ subgraph that connects to $a$, then it also connects to $b$, thus the two terms in the numerator cancel and $|\psi^\triangleright\rangle$ is indeed an eigenstate of $H_L$ with $0$ energy. 

Next we turn to the state
\begin{equation}
|\psi^{\diamond}_\sigma \rangle \equiv \frac{|a\rangle + \sigma  |b\rangle - |c\rangle -\sigma  |d\rangle}{2}
\end{equation}
with $\sigma = \pm 1$ and act again with $H_L$
\begin{equation}
H_L |\psi^{\diamond}_\sigma \rangle =  \sum_{i}  \frac{h_{ia}+ \sigma  h_{ib}-h_{ic} -h_{id}\sigma  }{2} |i\rangle .
\end{equation}
In the $\diamond $ subgraph if a vertex $i$ that is not $b$ is connected to $a$, it is also connected to $c$. Thus the first and third term cancel if $i\neq b,d$. Similarly the second and fourth terms cancel if $i\neq a,c$. Thus we are left with four terms corresponding to $i=a,b,c,d$:
\begin{equation}
H_L |\psi^{\diamond}_\sigma \rangle =  \frac{ \sigma }{2} |a\rangle + \frac{1}{2} |b\rangle -\frac{\sigma  }{2} |c\rangle - \frac{1   }{2} |d\rangle = \sigma  |\psi^{\diamond}_\sigma \rangle, 
\end{equation}
where we used that $h_{ii}=h_{ac}=h_{ad}=h_{bc}=h_{bd}=0$ and $h_{ab} = h_{cd} = 1$. Thus $|\psi^{\diamond}_\sigma \rangle$ is an eigenstate with energy $\sigma$.
\section{Probability distribution for the appearance of subgraphs with sparse eigenstates}
\label{app:probdist}
As discussed in the main text, the appearance of certain scars is tied to the appearance of either one of the special subgraphs, shown in Fig. \ref{Fig:Eigen}, within the adjacency graph of $H_L$. Here we determine the probability for the individual subgraphs as a function of the edge insertion probability $p_\text{edge}$ . We pick for the sub-Hilbert spaces a dimension $\mathcal D = 20$ and proceed by random sampling. We generate $1000$ random graphs for any given value of $p_\text{edge}$ and check if the special subgraphs appear therein. The frequency of appearance of the triangle subgraph is shown in Fig. \ref{FigSubGraphDistrTri} (a) and for the diamond subgraph in Fig. \ref{FigSubGraphDistrTri} (b). 
\begin{figure}[h] 
\centering{}\includegraphics[width= 0.9\columnwidth]{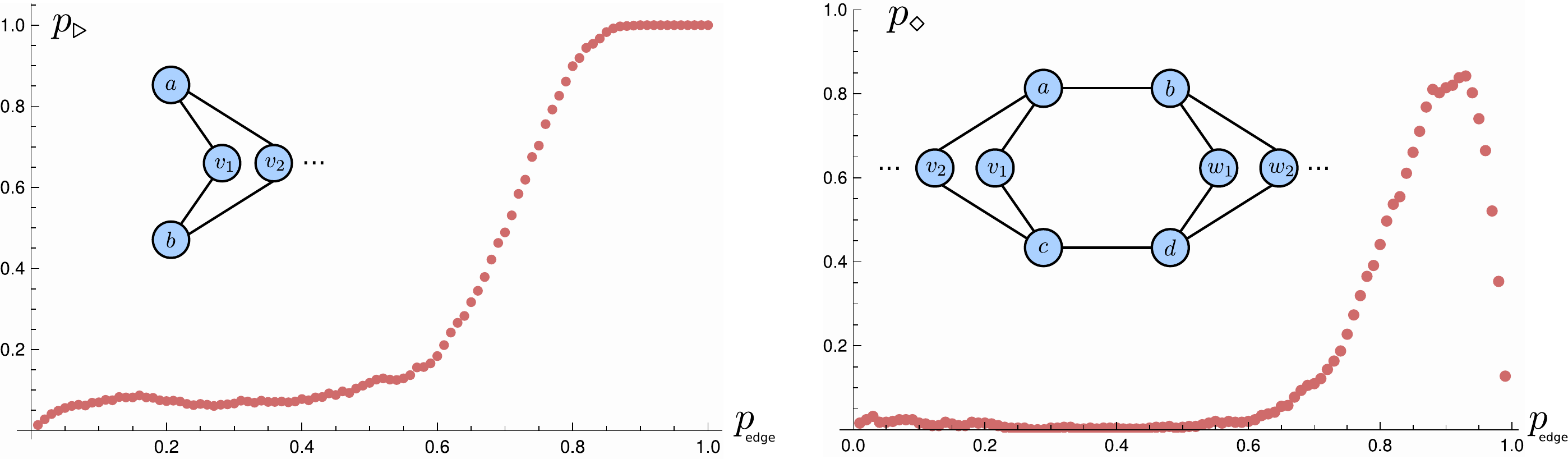}\protect\caption{\label{FigSubGraphDistrTri} Probability distributions for the occurence in $H_L$ of at least one subgraph of the shown type, in both cases $\mathcal D = 20$ (the value of $\mathcal D'$ is immaterial here, since the subgraph appears in $H_L$ and $H_R$). On the vertical is the probability of occurence and on the horizontal the edge probability $p_\text{edge}$ used in the construction of $H_L$.}
\end{figure}

\section{The PXP Hamiltonian in the Gray code basis}
\label{sec:GrayCode}

An explicit construction of a Hamiltonian-path through a Fibonacci cube was given in \cite{Zelina2006} and proceeds in a recursive way as follows. For $L=1$ we order the two basis states as ${\mathcal B_1 = \left(|\circ\rangle, |\bullet \rangle \right)}$ and for $L=2$ we use the ordering $\mathcal B_2 = (|\circ \bullet \rangle, |\circ\circ\rangle, |\bullet \circ \rangle)$.  Given basis state numberings  $\mathcal B_{L-2}$ and $\mathcal B_{L-1}$, we first construct the order-reversed tuples $\mathcal{ \overline  B}_{L-2}$ and $\mathcal{ \overline  B}_{L-1}$. From the latter we obtain $\mathcal B_{L}$ by concatenation of sites:
\begin{equation}
\mathcal B_L \equiv \left(|\circ\rangle \otimes \mathcal{ \overline  B}_{L-1}, |\bullet \circ\rangle \otimes  \mathcal{ \overline  B}_{L-2}  \right).
\label{eq:GrayCodeConstr}
\end{equation}
As an example, from $\mathcal{B}_1$ and $\mathcal{B}_2$ we obtain
\begin{equation}
\mathcal{B}_3 =(|{\circ \bullet \circ} \rangle, | {\circ \circ \circ} \rangle, | {\circ \circ\bullet}\rangle, |{\bullet \circ \bullet}\rangle, |{\bullet \circ \circ} \rangle),
\end{equation}
which is clearly a sequence that can be traversed by changing precisely one site at a time. Such an ordering is sometimes  called a Gray code and has many applications in computer science \cite{Press1988} and appears famously in the solution of Cardano's ring puzzle \cite{Gardner1972}.

In Fig. \ref{Fig1}  (a) the Hamiltonian path is shown for the $L=6$ PXP configurations. The corresponding adjacency matrix in the Gray code basis is shown in Fig. \ref{Fig1} (b). The super- and subdiagonals correspond to the Hamiltonian path. The skew-diagonals can be understood by considering an example. The basis states $|2\rangle$ and $|3\rangle$ are connected by the Hamiltonian for the left part of the system. Thus the states $|2\rangle|\bar i\rangle$ and $|3\rangle|\bar i\rangle$ are connected for all $i$ by the full Hamiltonian. These states correspond to the second and third rows in Fig. \ref{Fig1} (a). The states that are connected are labeled in the Gray code basis as $(6,15)$, $(7,14)$, $(8,13)$, $(9,12)$ and $(10,11)$. Thus there will be five entries $1$ in the adjacency matrix at position $(i,j)$, where $i+j = 21$. This corresponds to a skew diagonal.

\section{The PXP Hamiltonian in the Zeckendorf basis}
\label{sec:Zeckendorf}

There is a mathematical result, called the Zeckendorf theorem \cite{Zeckendorf1972, Graham1994a, Knuth1988}, which guarantees that every  integer $n\geq 0$ can be uniquely represented as a sum of Fibonacci numbers without making use of two consecutive Fibonacci numbers:
\begin{eqnarray}
n=\sum_{i=2}^{\infty}d_{i}F_{i}, \ \ \ d_i\in\{0,1\}, \ \ \ d_i d_{i+1}=0 .
\end{eqnarray}
The string $(d_2,d_3, \dots)$ defines a kind of binary expansion of the integer $n$. Since no two $1$'s are adjacent to each other in the expansion this corresponds to a valid PXP state if we identify $1 \leftrightarrow \bullet$ and $0 \leftrightarrow \circ$. Thus the Zeckendorf numbering provides a one-to-one mapping between valid PXP configurations and the integers in the interval $[0,F_{L+2}-1]$.  Thus we can label all of the valid PXP configurations of length $L$ by basis states $|0\rangle, |1\rangle,\dots,|F_{L+2}-1\rangle$
The systematics of the Zeckendorf numbering allows us to iteratively construct the Hamiltonian of size $L+1$ from the Hamiltonians of size $L$ and $L-1$. To construct a basis state for $H_{L+1}$, we can append an empty site $\circ$ to a basis state of size $L$. This does not change the index of the basis state, since in the Zeckendorf expansion this merely adds a $0$. The remaining basis states are obtained by appending $\circ \bullet$ to a basis state of size $L-1$. Then we obtain a basis state with index increased by $F_{L+2}$ according to the Zeckendorf expansion. In this basis we have the following recursive block-matrix relation

\begin{eqnarray}
H_{L+1}=\left(\begin{array}{cc}
H_{L} & P_{L}\\
P_{L}^{T} & H_{L-1}
\end{array}\right),
\label{eq:iterA}
\end{eqnarray}
where $P_L$ is a projection matrix of dimension $F_{L+2}\times F_{L+1}$ with entries $(P_L)_{ij} = \delta_{ij}$. We can generate the Hamiltonians for arbitrary sizes in an iterative way. The first few matrices are
\[
H_{2}=\left(\begin{array}{cc}
0 & 1\\
1 & 0
\end{array}\right)\ \ \ P_{2}=\left(\begin{array}{c}
1\\
0
\end{array}\right)
\]
\[
H_{3}=\left(\begin{array}{ccc}
0 & 1 & 1\\
1 & 0 & 0\\
1 & 0 & 0
\end{array}\right)\ \ \ P_{3}=\left(\begin{array}{cc}
1 & 0\\
0 & 1\\
0 & 0
\end{array}\right)
\]
\begin{figure}[b] 
\centering{}\includegraphics[width= 0.4\columnwidth]{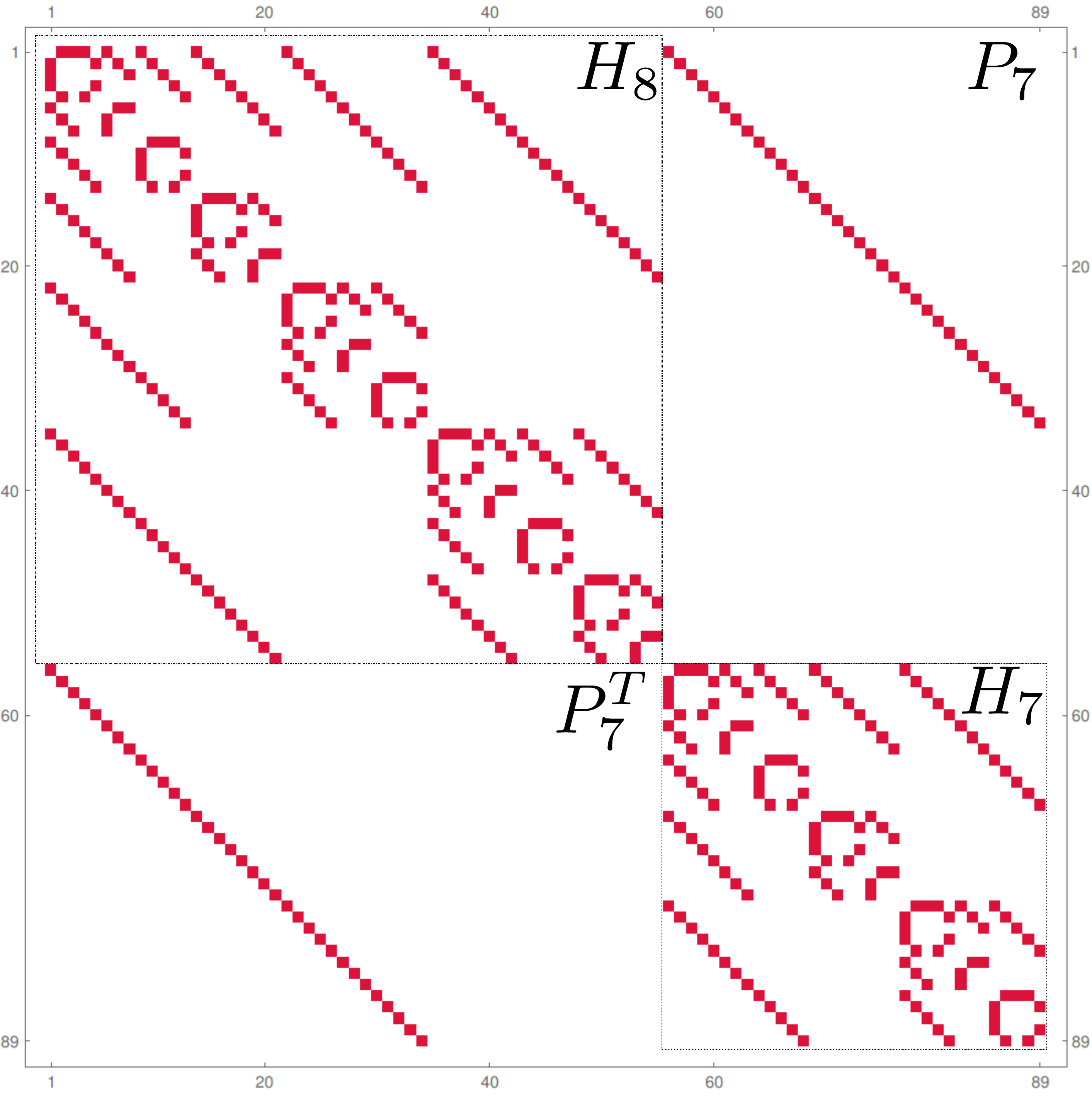}\protect\caption{\label{Fig:AdjMatrixH9} In the Zeckendorf basis the $89 \times 89$ adjacency matrix of $H_{9}$ takes on a simple nearly self-similar form.}
\end{figure}
\[
H_{4}=\left(\begin{array}{ccccc}
0 & 1 & 1 & 1 & 0\\
1 & 0 & 0 & 0 & 1\\
1 & 0 & 0 & 0 & 0\\
1 & 0 & 0 & 0 & 1\\
0 & 1 & 0 & 1 & 0
\end{array}\right)\ \ P_{4}=\left(\begin{array}{ccc}
1 & 0 & 0\\
0 & 1 & 0\\
0 & 0 & 1\\
0 & 0 & 0\\
0 & 0 & 0
\end{array}\right).
\]
The Fig. \ref{Fig:AdjMatrixH9} shows $H_9$ as an example and illustrates the near self-similar quality of the matrices. 
It is illuminating to rewrite the Hamiltonian $H_{L+1}$ in a slightly different form:
\begin{eqnarray}
H_{L+1}&=& \Pi^T_{L}
\left(\begin{array}{cc}
H_{L} &\mathbf{1} \\
\mathbf{1}  & H_{L} 
\end{array}\right)\Pi_{L} = \Pi^T_{L} \left(H_L\square K_2\right) \Pi_{L}
\label{eq:iterArecast}
\\
\Pi_{L} &\equiv& \left(\begin{array}{cc}
\mathbf{1} & {\mathbf 0}\\
{\mathbf 0}& P_L
\end{array}\right)
\label{eq:iterPidef}
\end{eqnarray}
The factor in the center of eq. \eqref{eq:iterArecast} has a special meaning. If we denote the graphs corresponding to the adjacency matrices $H_L$ by $G_L$, then the factor in the center is known in the graph theory literature as the adjacency matrix of the cartesian product of $G_L$ and $K_2$, where $K_2$ is the complete graph on $2$ vertices. This product describes a graph that is obtained from $G_L$ by duplicating and connecting the identical vertices by edges. On this the $\Pi_{L}$ operator acts and projects out the PXP-constraint-violating vertices. If we had not included the projection $\Pi_{L}$, we would have ended up with the hyper-cube graphs $Q_{L}$, since these are constructed recursively as $Q_{L+1} = Q_L \square K_2$. The presence of the projection operators $\Pi_{L}$ instead results in the Fibonacci-cube graph. It is therefore not surprising that repeating this cycle of duplication followed by projection yields nearly self-similar matrices $H_L$.

The adjaceny graphs of $H_L$ are all bipartite, in other words one can divide the Fibonacci-cube into two sublattices of $A$- and $B$-type such that edges only connect vertices of $A$ to vertices of $B$. The underlying reason for this is that the number of excited sites $N_\bullet$ in a configuration differs by $\pm 1$ from a neighboring configuration. Hence the value of $N_\bullet \mod 2$  can be used to assign a vertex uniquely to a sublattice. 

Next we consider an eigenvector $\psi$ of $H_L$. We introduce an operator $Q_L$ that changes the sign of the components of $\psi$ on the $B$ sublattice. This operator can be constructed iteratively similar to eq. \eqref{eq:iterA}:
\begin{eqnarray}
Q_{L+1}=\left(\begin{array}{cc}
Q_{L} & \bm 0\\
\bm 0 & -Q_{L-1}
\end{array}\right),
\label{eq:iterQ}
\end{eqnarray}
where $\bm 0$ are rectangular zero-matrices. The initial matrices are $Q_1 = \text{diag}(1,-1)$ and $Q_2=\text{diag}(1,-1,-1)$. From the discussion in App. \ref{app:Bip} it follows that if $\psi$ is an eigenstate of $H_L$ with energy $E$, then $Q_L \psi$ is an eigenstate with energy $-E$ and the spectrum is symmetric. 

\section{Level statistics analysis and non-integrability of the model}
\label{sec:LevelStat}
To demonstrate that the Hamiltonians $H$ of eq. \eqref{eq:DefHamiltonian} are in general quantum-ergodic, we carry out a level statistics analysis. We generate $1000$ random graphs for $H$ from the ensemble $\mathcal G(20,14,0.4)$. The resulting Hamiltonians have dimension $20^2 - (20-14)^2 = 364$.  Next we compute for each $H$ all its energy levels $E_i$, remove one of the inversion-symmetry sectors and sort the energies in ascending order $E_i \leq E_{i+1}$. Following \cite{Oganesyan2007} we compute the ratio of successive energy differences $\delta E_i \equiv E_{i+1} - E_i$, defined as $r_i ={\min(\delta E_i ,  \delta E_{i-1})}/{\max(\delta E_i ,  \delta E_{i-1})}$ and generate a histogram.
When the system is non-integrable the energy-levels repel and show Wigner-Dyson level statistics, while for integrable models the levels may cross and therefore give rise to Poisson distributed level-spacings. The distribution for the ratios $r_i$ was worked out in \cite{Atas2013} in the spirit of Wigner's surmise \cite{Wigner1957} and is 
\begin{equation}
\label{eq:WDProb}
P(r) = \frac{27}{4} \frac{r+r^2}{(1+r+r^2)^{5/2}}
\end{equation}
for the Wigner-Dyson case. 
\begin{figure}[b!]
\centering{}\includegraphics[width= 0.5\columnwidth]{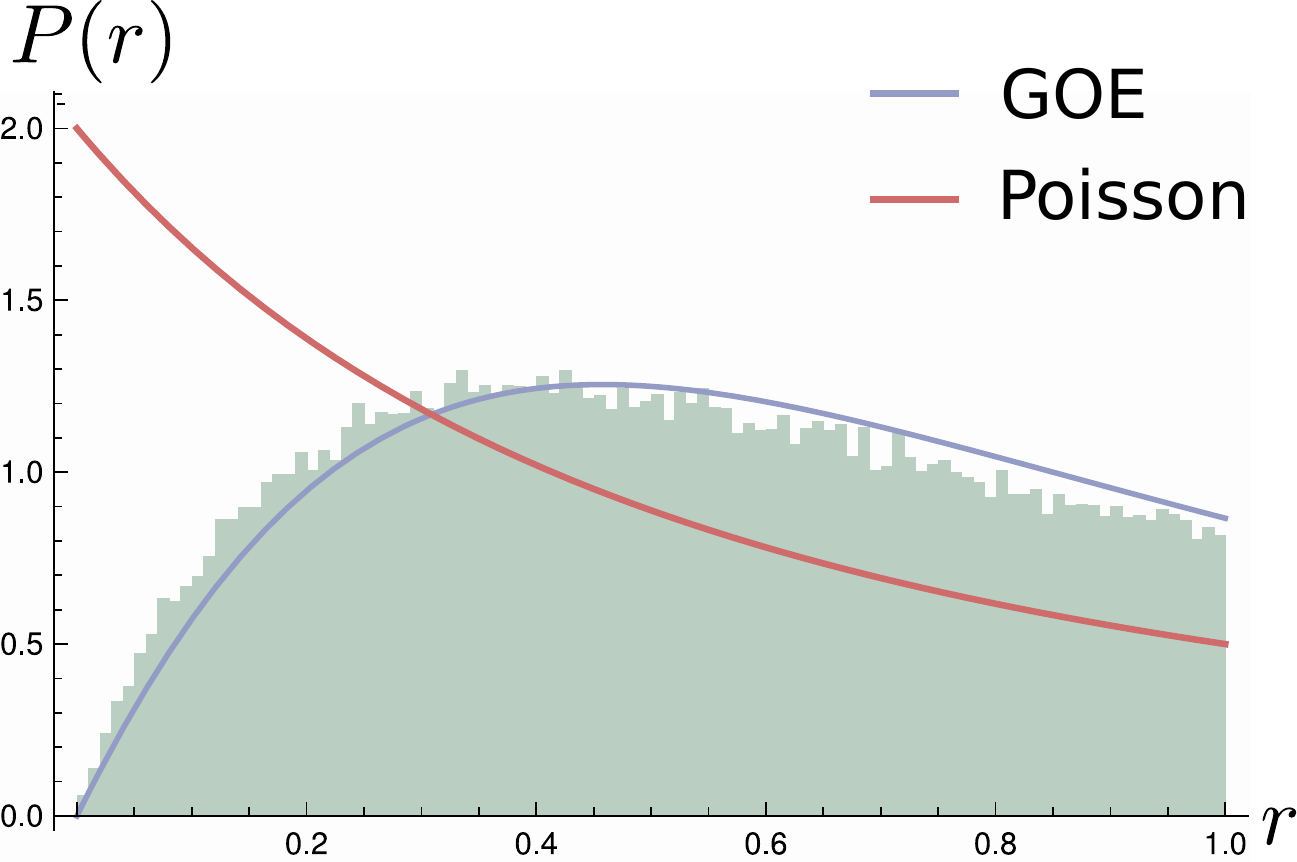}\caption{\label{Fig:LevelStat} Distribution of ratios of consecutive level-spacings for the eigenstates of Hamiltonians $H$ with $+1$ inversion quantum number from the $\mathcal G(20,14,0.4)$ ensemble. The ratio $r$ follows the probability density in eq. \eqref{eq:WDProb} for matrices in the Gaussian orthogonal ensemble, implying a non-integrable Hamiltonian $H$. This is also confirmed by comparing our numerical average $\langle r \rangle \approx 0.517$ to the exact mean ${\langle r \rangle_{\text{GOE}} = 4-2\sqrt{3}\approx0.536}$.}
\end{figure}
The corresponding distribution for integrable systems  with an underlying Poisson statistics follows the distribution
\begin{equation}
\label{eq:Poiss}
P(r) = \frac{2}{(1+r)^2}.
\end{equation}
Our numerical result in Fig. \ref{Fig:LevelStat} agrees well with Wigner-Dyson statistics eq. \eqref{eq:WDProb}. We have found similar behavior for other generic values of $(\mathcal D, \mathcal D' ,p)$, leading us to conclusion that the model is non-integrable in general. The exception is the non-interacting point $\mathcal D = \mathcal D'$, where we find Poisson statistics for generic values of $p$.

\section{The maximum overlap between eigenstates of the constrained and unconstrained Hamiltonian}
\label{app:overlap}
In this section we generate $H_L$ from the $\mathcal G(\mathcal D, \mathcal D', p)$ ensemble as explained in the main text  and construct from this the unconstrained Hamiltonian
\begin{equation}
\tilde H =H_L \otimes \mathbb{I}_R  + \mathbb{I}_L \otimes  H_R.
\end{equation}
We obtain $H$ of eq. \eqref{eq:DefHamiltonian} by nulling the rows and columns of $\tilde H$ that contain the incompatible basis states. In this way $H$ and $\tilde H$ retain the same dimension. Next we compute the normalized eigenstates $|\psi_i\rangle$ of $H$ and $|\varphi_j\rangle$ of $\tilde H$. We are interested in the maximum overlap $|\langle \psi_i | \varphi_j\rangle|^2$ over all $1\leq i,j\leq \mathcal D$. We fix $\mathcal D = 20$ and for each value of $\mathcal D'$ and $p$ we generate around $100$ matrices $H$ and $\tilde H$, compute for each case the largest overlap square and take the ensemble average. 
\begin{figure*}[t] 
\centering{}\includegraphics[width=0.55\columnwidth]{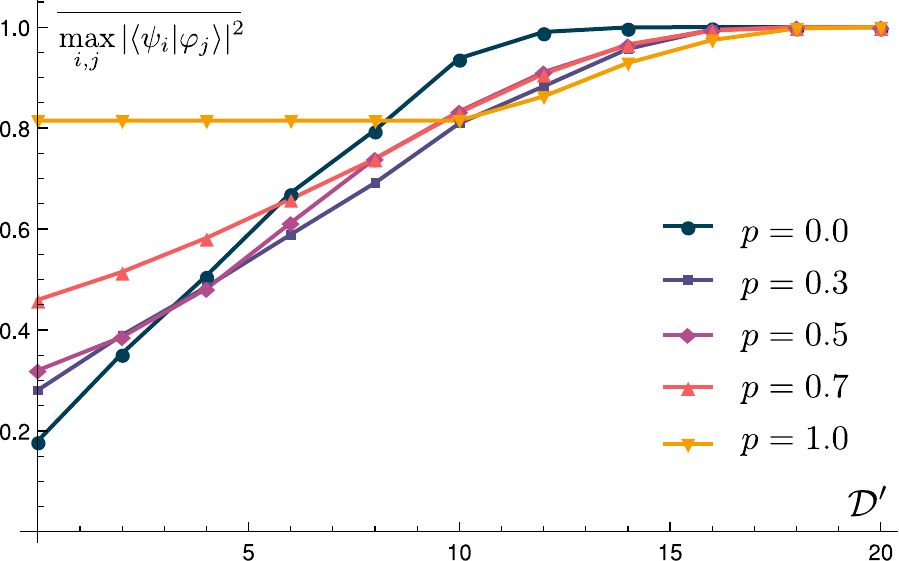}\caption{\label{Fig:Maxoverlap} The ensemble average of the maximum overlap between the normalized eigenstates of $H$ and $\tilde H$ as a function of the cutoff $\mathcal D'$ in $\mathcal G(20, \mathcal D', p)$. }
\end{figure*}
The result is shown in Fig. \ref{Fig:Maxoverlap} as a function of $\mathcal D'$ for several values of $p$. Clearly among the eigenstates of $H$ we seem to always find at least one eigenvector that has a large overlap with an eigenvector of $\tilde H$. 

As discussed in the main text, this is ultimately the reason for the appearance of the boxed scars. We cannot give a proof for the existence of such an eigenvector, but indicate heuristically the mechanism for it appearance. Let us take a real symmetric matrix $\mathcal D \times \mathcal D $ matrix $H$ and consider removing the last row $\bm v^T$ and column $\bm  v$. We consider the eigenvalue problem for the matrix $H$:
\begin{equation}
\label{eq:blockmatrixHeuristic}
\left(\begin{array}{cc}
H_0 & \bm{v}\\
\bm{v}^{T} & 0
\end{array}\right)\left(\begin{array}{c}
\bm{\psi}\\
\psi_{0}
\end{array}\right)=\lambda\left(\begin{array}{c}
\bm{\psi}\\
\psi_{0}
\end{array}\right),
\end{equation}
where $\bm \psi$ is a $\mathcal D-1$ dimensional real vector, $\psi_0$ is a real number and $H_0$ is a $(\mathcal D -1)\times (\mathcal D-1) $ block matrix. This yields the equation $\lambda \psi_0 = \bm v \cdot \bm \psi$. Substituting this back, we obtain
\begin{equation}
H_0\bm{\psi}+\frac{1}{\lambda}(\bm{v}\cdot\bm{\psi})\bm{v}=\lambda\bm{\psi}.
\end{equation}
In general, the vector $\bm \psi$ will not be an eigenvector of $H_0$, since the second term on the left hand side spoils the eigenvector property. Yet, it is well-known that two random vectors in a high-dimensional space will, with very high probability, be nearly orthogonal to each other. Thus there is a good chance that for some of the eigenvectors of $H$ the corresponding $\bm \psi$ vectors are orthogonal to $\bm v$, rendering the second term negligible. Such vectors $\bm  \psi$ would then be close to being eigenvectors of $H_0$. As we continue to reduce the matrix by removing rows and columns by the same process, eventually $\bm \psi$ may not be orthogonal to the removed $\bm v$ and therefore drop out as a candidate eigenvector. Such a picture is consistent with what is observed in Fig. \ref{Fig:Maxoverlap}. 

\section{The quantum dimer ladder and the exact entanglement entropy in the RK state}
\label{app:qdmladder}
\begin{figure}[b] 
\centering{}\includegraphics[width= 5.5 cm]{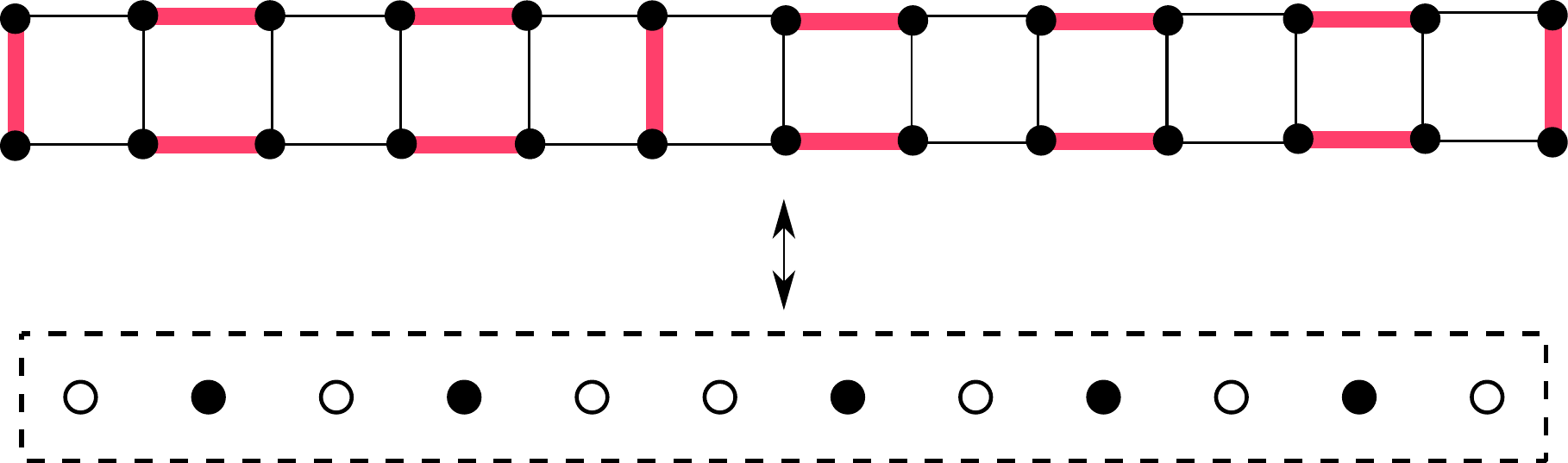}\protect\caption{\label{FigMapping} As noticed by \cite{Chepiga2019} there is an equivalence between the $2\times L$ QDM and the PXP model of size $L-1$ . \label{fig:Intro} }
\end{figure}
The quantum dimer ladder with $L$ plaquettes is shown in Fig. \ref{FigMapping}. Its dynamics is determined by the Rokhsar-Kivelson Hamiltonian
\begin{equation}
H_{\text{RK}}=\sum_\square \left|\vcenter{\hbox{\includegraphics[height=0.016\textheight]{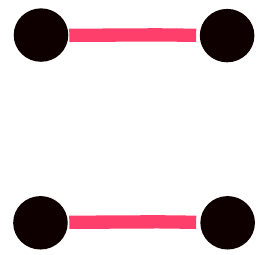}}}\right\rangle \left\langle \vcenter{\hbox{\includegraphics[height=0.016\textheight]{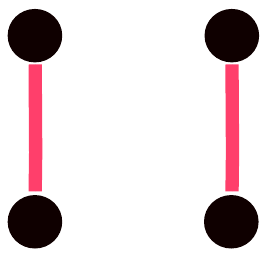}}}\right|+\left|\vcenter{\hbox{\includegraphics[height=0.016\textheight]{terms/term1.pdf}}}\right\rangle \left\langle \vcenter{\hbox{\includegraphics[height=0.016\textheight]{terms/term2.pdf}}}\right|,
\end{equation}
where the sum is over all $L$ plaquettes.
 It was noticed by \cite{Chepiga2019} that this model is equivalent to the PXP model. This becomes manifest once one identifies the plaquettes with horizontal dimer pairs by an occupied site $\bullet$ and all other plaquette configurations, of which there are four, with an empty site $\circ$ .  The PXP constraint is equivalent to the `one dimer per site' condition. The latter implies that two neighboring plaquettes can never both be occupied by horizontal dimer pairs.

In quantum dimer models it is natural to have a potential energy term that counts the number of flippable plaquettes. 
\begin{equation}
H_{\text{pot}}=V_{\text{pot}}\sum_\square \left|\vcenter{\hbox{\includegraphics[height=0.016\textheight]{terms/term2.pdf}}}\right\rangle \left\langle \vcenter{\hbox{\includegraphics[height=0.016\textheight]{terms/term2.pdf}}}\right|+\left|\vcenter{\hbox{\includegraphics[height=0.016\textheight]{terms/term1.pdf}}}\right\rangle \left\langle \vcenter{\hbox{\includegraphics[height=0.016\textheight]{terms/term1.pdf}}}\right|,
\end{equation}
The full Hamiltonian is now 
\begin{equation}
H=H_\text{RK} + H_\text{pot}
\end{equation}
As an aside, we note that while it is natural to have this term for a quantum dimer model, in the language of the PXP model this takes on the following expression 
\[
H_{\text{pot}}/V_{\text{pot}}=\frac{1-Z_{2}}{2}+\sum_{i=1}^{L-2}\frac{1-Z_{i-1}}{2}\frac{1-Z_{i+1}}{2}+\frac{1-Z_{L-1}}{2}.
\]
It is interesting to note that at $ V_\text{pot}=-1$ the Hamiltonian is proportional to the Laplacian matrix familiar in graph theory (this remark applies also to the original Rokhsar-Kivelson model). With our chosen sign conventions, the maximum energy state is an equal amplitude superposition of all basis states $|s\rangle$, while the ground state is an equal amplitude superposition with weights $W$ that are $+1$ for basis states of A-type (even number of horizontal dimer pairs) and $-1$ for B-type:
\begin{eqnarray}
|\tilde \psi_\text{RK}\rangle =\frac{1}{\sqrt{F_{L+1}}} \sum_s W_s |s\rangle
\end{eqnarray}

As shown in App. \ref{app:QEntanglement}, it has the same entanglement entropy as the state
\begin{eqnarray}
|\psi_\text{RK}\rangle  \equiv Q |\tilde \psi_\text{RK}\rangle= \frac{1}{\sqrt{F_{L+1}}} \sum_s  |s\rangle.
\end{eqnarray}
In the remainder of this section we compute the entanglement entropy of this state. It has the decomposition
\begin{align*}
|\psi_{\text{RK}}\rangle & =\frac{1}{\sqrt{F_{L+1}}}\sum_{m,n}|m\rangle|n\rangle =\frac{1}{\sqrt{F_{L+1}}}\sum_{\alpha.\beta}|\alpha\ \vcenter{\hbox{\includegraphics[height=0.016\textheight]{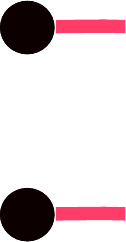}}}\rangle|\vcenter{\hbox{\includegraphics[height=0.016\textheight]{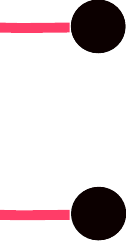}}}\ \beta\rangle+\frac{1}{\sqrt{F_{L+1}}}\sum_{\gamma.\delta}|\gamma\rangle|\delta\rangle,
\end{align*}
where $\alpha,\beta$ are dimer configuration strings of length $L/2-1$ and $\gamma,\delta$ are dimer configuration of length $L/2$. 

The density matrix of this state is 
\[
\rho=|\psi_{\text{RK}}\rangle\langle\psi_{\text{RK}}|.
\]
Tracing out the right part of the sytem, we obtain
\begin{align*}
\rho_{A}= & \frac{1}{F_{L+1}}\sum_{\alpha,\alpha'}N_{\beta}|\alpha\ \vcenter{\hbox{\includegraphics[height=0.016\textheight]{terms/termCUT1.pdf}}}\rangle\langle\alpha'\ \vcenter{\hbox{\includegraphics[height=0.016\textheight]{terms/termCUT1.pdf}}}|+\frac{1}{F_{L+1}}\sum_{\gamma,\gamma'}N_{\delta}|\gamma\rangle\langle\gamma'|.
\end{align*}
Here $N_{\beta},N_{\delta}$ are the numbers of different configurations
of $\beta$ and $\delta$. Since a configuration $\beta$ has length
$L/2-1$ and $\delta$ has length $L/2$ we have 
\begin{align*}
N_{\beta} & =F_{L/2}\\
N_{\delta} & =F_{L/2+1}
\end{align*}
Thus we have a very simple block form for the reduced density matrix
$\rho_{A}$:
\[
\rho_{A}=\frac{1}{F_{L+1}}\left(\begin{array}{cc}
\rho_{A}^{(\alpha)} & 0\\
0 & \rho_{A}^{(\gamma)}
\end{array}\right),
\]
where the $\rho_{A}^{(i)}$ are matrices with all entries equal:
\begin{align*}
\rho_{A}^{(\alpha)} & =\left(\begin{array}{ccc}
F_{L/2} & \dots & F_{L/2}\\
\vdots & \ddots & \vdots\\
F_{L/2} & \dots & F_{L/2}
\end{array}\right)\\
\rho_{A}^{(\gamma)} & =\left(\begin{array}{ccc}
F_{L/2+1} & \dots & F_{L/2+1}\\
\vdots & \ddots & \vdots\\
F_{L/2+1} & \dots & F_{L/2+1}
\end{array}\right).
\end{align*}
Here the dimension of $\rho_{A}^{(\alpha)}$ is $F_{L/2}\times F_{L/2}$ and that of $\rho_{A}^{(\gamma)}$ is $F_{L/2+1}\times F_{L/2+1}$.
It is straightforward to find the eigenvalues of $\rho_{A}$:
\[
\frac{F_{L/2}^{2}}{F_{L+1}},\frac{F_{L/2+1}^{2}}{F_{L+1}},0,\dots0,.
\]
i.e. only two eigenvalues are non-zero. Note that the trace of the reduced density matrix is $1$ by virtue of the Fibonacci identity
\[
{F_{n}^{2}}+F_{n+1}^{2} = F_{2n+1}
\]
Finally we compute from this the entanglement entropy of the ground state of a size $L$ system at the RK point:
\[
S_{\text{ent}}(L)=-\frac{F_{{L}/{2}}^{2}}{F_{L+1}}\log\frac{F_{{L}/{2}}^{2}}{F_{L+1}}-\frac{F_{{L}/{2}+1}^{2}}{F_{L+1}}\log\frac{F_{{L}/{2}+1}^{2}}{F_{L+1}}
\]
This formula agrees exactly with the numerical result. For large $L$ we can use the asymptotic formula $F_{n} \sim\frac{\phi^{n}}{\sqrt{5}}$ with  $\phi =\frac{1+\sqrt{5}}{2}$
to derive the relation
\begin{eqnarray}
S_{\text{ent}}(L\rightarrow\infty)&=&\frac{\log5+2\log\frac{1+\sqrt{5}}{2}-\left(3+\sqrt{5}\right)\log\frac{1+\sqrt{5}}{2\sqrt{5}}}{5+\sqrt{5}}\nonumber  \\
&\approx&0.589...\nonumber 
\end{eqnarray}
\end{widetext}
\bibliography{biblio}
\end{document}